\pdfoutput=1

\documentclass[prd,amsmath,superscriptaddress,%
               nobibnotes,nofootinbib,preprintnumbers,
               twocolumn]{revtex4-1}

\usepackage{graphicx}

\usepackage{ifthen}

\usepackage{verbatim}

\graphicspath{{figures/}}

\usepackage{listings}

\usepackage[usenames]{color}
\definecolor{lstbkgdcolor}{gray}{0.85}

\newlength{\snippetindent}
\newlength{\snippetskip}
\lstdefinestyle{Pstyle}{language=sh,
        xleftmargin=\snippetindent,xrightmargin=\snippetindent,
        aboveskip=\snippetskip, belowskip=\snippetskip,
        columns=fixed,basicstyle=\ttfamily,basewidth=0.5em,
        frame=single,
        backgroundcolor=\color{lstbkgdcolor},
        gobble=1
        }

\lstdefinestyle{Fstyle}{language=[95]Fortran,
        xleftmargin=\snippetindent,xrightmargin=\snippetindent,
        aboveskip=\snippetskip, belowskip=\snippetskip,
        columns=fixed,basicstyle=\ttfamily,basewidth=0.5em,
        frame=single,
        backgroundcolor=\color{lstbkgdcolor},
        gobble=1
        }

\lstdefinestyle{Cstyle}{language=C++,
        xleftmargin=\snippetindent,xrightmargin=\snippetindent,
        aboveskip=\snippetskip, belowskip=\snippetskip,
        columns=fixed,basicstyle=\ttfamily,basewidth=0.5em,
        frame=single,
        backgroundcolor=\color{lstbkgdcolor},
        gobble=1
        }

\lstnewenvironment{snippet}{\pagebreak[2]}{}
\lstnewenvironment{Psnippet}{\pagebreak[2]\lstset{style=Pstyle}}{\pagebreak[2]}
\lstnewenvironment{Fsnippet}{\pagebreak[2]\lstset{style=Fstyle}}{\pagebreak[2]}
\lstnewenvironment{Csnippet}{\pagebreak[2]\lstset{style=Cstyle}}{\pagebreak[2]}

\newcommand{\ie}{i.e.}
\newcommand{\eg}{e.g.}

\newcommand{\refeqn}[2][eqn:]{Eqn.~(\ref{#1#2})}
\newcommand{\Refeqn}[2][eqn:]{Equation~(\ref{#1#2})}
\newcommand{\reftab}[2][tab:]{Table~\ref{#1#2}}

\newcommand{\reffig}[2][fig:]{Figure~\ref{#1#2}}
\newcommand{\Reffig}[2][fig:]{Figure~\ref{#1#2}}
\newcommand{\refsec}[2][sec:]{Section~\ref{#1#2}} 
\newcommand{\Refsec}[2][sec:]{Section~\ref{#1#2}} 
\newcommand{\refapp}[2][sec:]{the Appendix}
\newcommand{\Refapp}[2][sec:]{The Appendix}

\newcommand{\ifmulticol}[2]{%
  \ifthenelse{\lengthtest{1.9\columnwidth<\textwidth}}{#1}{#2}%
}

\newcommand{\insertfig}[2][]{%
  \hspace*{\stretch{1}}
  \ifthenelse{\NOT\equal{#1}{}}%
    {\includegraphics[keepaspectratio,width=#1\columnwidth]{#2}}%
    {\ifthenelse{\lengthtest{1.9\columnwidth<\textwidth}}%
      {\includegraphics[keepaspectratio,width=1.05\columnwidth]{#2}}%
      {\includegraphics[keepaspectratio,width=0.70\columnwidth]{#2}}%
    }%
  \hspace*{\stretch{1}}
}

\newcommand{\insertwidefig}[2][]{%
  \hspace*{\stretch{1}}
  \ifthenelse{\NOT\equal{#1}{}}%
    {\includegraphics[keepaspectratio,width=#1\textwidth]{#2}}%
    {\ifthenelse{\lengthtest{1.9\columnwidth<\textwidth}}%
      {\includegraphics[keepaspectratio,width=0.80\textwidth]{#2}}%
      {\includegraphics[keepaspectratio,width=0.99\textwidth]{#2}}%
    }%
  \hspace*{\stretch{1}}
}

\newcommand{\orderof}[1]{\ensuremath{\mathcal{O}(#1)}}
\newcommand{\erf}{\mathop{\mathrm{erf}}}

\newcommand{\Enr}{\ensuremath{E}}

\newcommand{\dRdE}{\ensuremath{\frac{dR}{dE}}}
\newcommand{\dRdEnr}{\ensuremath{\frac{dR}{dE}}}

\newcommand{\Emin}{\ensuremath{E_{\mathrm{min}}}}

\newcommand{\mchi}{\ensuremath{m_{\chi}}}
\newcommand{\rhochi}{\ensuremath{\rho_{0}}}
\newcommand{\nchi}{\ensuremath{n_{0}}}

\newcommand{\vmin}{\ensuremath{v_\mathrm{min}}}
\newcommand{\vmp}{\ensuremath{v_0}}
\newcommand{\vrot}{\ensuremath{v_\mathrm{rot}}}

\newcommand{\vesc}{\ensuremath{v_\mathrm{esc}}}
\newcommand{\Nesc}{\ensuremath{N_\mathrm{esc}}}
\newcommand{\bv}{\ensuremath{\mathbf{v}}}  
\newcommand{\bvsun}{\ensuremath{\mathbf{v}_{\odot}}}
\newcommand{\bvsunpec}{\ensuremath{\mathbf{v}_{\odot,\mathrm{pec}}}}
\newcommand{\bvLSR}{\ensuremath{\mathbf{v}_{\mathrm{LSR}}}}

\newcommand{\qmax}{\ensuremath{q_{\mathrm{max}}}}  

\newcommand{\fNSI}{\ensuremath{f_{N}}}
\newcommand{\fpSI}{\ensuremath{f_{\mathrm{p}}}}
\newcommand{\fnSI}{\ensuremath{f_{\mathrm{n}}}}

\newcommand{\aNSD}{\ensuremath{a_{N}}}
\newcommand{\apSD}{\ensuremath{a_{\mathrm{p}}}}
\newcommand{\anSD}{\ensuremath{a_{\mathrm{n}}}}

\newcommand{\GNSI}{\ensuremath{G_{\scriptscriptstyle\mathrm{SI}}^{N}}}
\newcommand{\GpSI}{\ensuremath{G_{\scriptscriptstyle\mathrm{SI}}^{\mathrm{p}}}}
\newcommand{\GnSI}{\ensuremath{G_{\scriptscriptstyle\mathrm{SI}}^{\mathrm{n}}}}

\newcommand{\GNSD}{\ensuremath{G_{\scriptscriptstyle\mathrm{SD}}^{N}}}
\newcommand{\GpSD}{\ensuremath{G_{\scriptscriptstyle\mathrm{SD}}^{\mathrm{p}}}}
\newcommand{\GnSD}{\ensuremath{G_{\scriptscriptstyle\mathrm{SD}}^{\mathrm{n}}}}

\newcommand{\sigmaSI}{\ensuremath{\sigma_{\mathrm{SI}}}}
\newcommand{\sigmaSD}{\ensuremath{\sigma_{\mathrm{SD}}}}

\newcommand{\sigmapSI}{\ensuremath{\sigma_{\mathrm{p,SI}}}}
\newcommand{\sigmanSI}{\ensuremath{\sigma_{\mathrm{n,SI}}}}

\newcommand{\sigmapSD}{\ensuremath{\sigma_{\mathrm{p,SD}}}}
\newcommand{\sigmanSD}{\ensuremath{\sigma_{\mathrm{n,SD}}}}
\newcommand{\sigmas}{\ensuremath{\boldsymbol{\sigma}}}

\newcommand{\gxa}{\ensuremath{g_{\chi a}}}
\newcommand{\gfa}{\ensuremath{g_{fa}}}
\newcommand{\effaN}{\ensuremath{\tilde{a}_{N}}}
\newcommand{\effap}{\ensuremath{\tilde{a}_{\mathrm{p}}}}
\newcommand{\effan}{\ensuremath{\tilde{a}_{\mathrm{n}}}}
\newcommand{\effsigma}{\ensuremath{\sigma_{(a,a)}}}
\newcommand{\effsigmap}{\ensuremath{\sigma_{\chi\mathrm{p}}}}
\newcommand{\effsigman}{\ensuremath{\sigma_{\chi\mathrm{n}}}}

\newcommand{\mup}{\ensuremath{\mu_{\mathrm{p}}}}

\newcommand{\Sp}{\ensuremath{\langle S_{\mathrm{p}} \rangle}}
\newcommand{\Sn}{\ensuremath{\langle S_{\mathrm{n}} \rangle}}
\newcommand{\Spp}{\ensuremath{S_{\mathrm{pp}}}}
\newcommand{\Spn}{\ensuremath{S_{\mathrm{pn}}}}
\newcommand{\Snn}{\ensuremath{S_{\mathrm{nn}}}}

\newcommand{\Cpp}{%
  C\nolinebreak\hspace{-.05em}\raisebox{.4ex}{\tiny\bf +}%
   \nolinebreak\hspace{-.10em}\raisebox{.4ex}{\tiny\bf +}}

\newcommand{\LUXCalc}{\texttt{LUXCalc}}
\newcommand{\TPCMC}{\texttt{TPCMC}}
\newcommand{\NEST}{\texttt{NEST}}
\newcommand{\DarkSUSY}{\texttt{DarkSUSY}}
\newcommand{\micrOMEGAs}{\texttt{micrOMEGAs}}  
\newcommand{\MicrOMEGAs}{\texttt{MicrOMEGAs}}  

\usepackage{hyperref}

\begin{document}


\title{LUX likelihood and limits on spin-independent and spin-dependent WIMP couplings with LUXCalc}



\author{Christopher Savage}
\email[]{chris@savage.name}
\affiliation{
  Department of Physics \& Astronomy,
  University of Utah,
  Salt Lake City, UT 84112}
\affiliation{
  Nordita,
  KTH Royal Institute of Technology and Stockholm University,
  SE-106 91 Stockholm, Sweden}
 
\author{Andre Scaffidi}
\email[]{andre.scaffidi@adelaide.edu.au}
\affiliation{
  ARC Center of Excellence for Particle Physics at the Terascale \& CSSM,
  Department of Physics,
  University of Adelaide}

\author{Martin White}
\email[]{martin.white@adelaide.edu.au}
\affiliation{
  ARC Center of Excellence for Particle Physics at the Terascale \& CSSM,
  Department of Physics,
  University of Adelaide}

\author{Anthony G. Williams}
\email[]{anthony.williams@adelaide.edu.au}
\affiliation{
  ARC Center of Excellence for Particle Physics at the Terascale \& CSSM,
  Department of Physics,
  University of Adelaide}

\date{\today}
 


\preprint{ADP-15-7/T909}
\preprint{NORDITA-2015-15}

\begin{abstract} 


We present \LUXCalc{}, a new utility for calculating likelihoods and deriving WIMP-nucleon coupling limits from the recent results of the LUX direct search dark matter experiment. After a brief review of WIMP-nucleon scattering, we derive LUX limits on the spin-dependent WIMP-nucleon couplings over a broad range of WIMP masses, under standard assumptions on the relevant astrophysical parameters. We find that, under these and other common assumptions, LUX excludes the entire spin-dependent parameter space consistent with a dark matter interpretation of DAMA's anomalous signal, the first time a single experiment has been able to do so. We also revisit the case of spin-independent couplings, and demonstrate good agreement between our results and the published LUX results. Finally, we derive constraints on the parameters of an effective dark matter theory in which a spin-1 mediator interacts with a fermionic WIMP and Standard Model fermions via axial-vector couplings. A detailed appendix describes the use of \LUXCalc{} with standard codes to place constraints on generic dark matter theories.

\end{abstract} 

\maketitle



\section{Introduction}
\label{sec:Intro}

Evidence for a large amount of non-baryonic ``dark matter'' (DM) in the universe has been accumulating for decades \cite{Zwicky:1937zza,Bertone:2004pz}. Recent observations of the cosmic microwave background have provided a precise measurement of the DM relic density, and also strongly support the idea of ``cold'' DM in the form of weakly interacting massive particles (WIMPs) \cite{Adam:2015rua,Planck:2015xua,Komatsu:2010fb}. The failure of the Standard Model (SM) of particle physics to adequately explain a variety of astrophysical observations has prompted the development of a large number of particle theories beyond the SM \cite{Bergstrom:2000pn}.

Concurrently with these theoretical developments, a large number of experiments have been conducted to search for DM annihilation in distant astrophysical objects, produce and observe DM particles in high energy particle collisions, or observe the direct interaction of particles of DM with Earth bound detectors. Although there are tantalising hints of DM signatures in one or more of these experiments, there is as yet no uncontroversial detection of (non-gravitational) WIMP interactions with ordinary matter \cite{Beringer:1900zz}. Given a particular new theory of particle physics, it is sometimes challenging to assess the likelihood of the model (as a function of the model parameters) given the null results of these experiments.

In this paper, we present \LUXCalc{}, a new utility for assessing the likelihood of new physics models given the recent results of the LUX experiment \cite{Akerib:2013tjd}, the most constraining direct search experiment to date for a wide range of WIMP models. In addition to describing the use of the tool for the general user, we apply it to derive LUX limits on spin-independent (SI) and spin-dependent (SD) WIMP-nucleon couplings. The former show good agreement with the official LUX results, thus validating our approach. Whilst SD limits have not been provided by the LUX collaboration, our results in the neutron-only coupling case are in close agreement with those calculated by Ref.~\cite{Buchmueller:2014yoa}.\footnote{Reference~\cite{Gresham:2013mua} first produced SD LUX limits, though only for a narrow range of WIMP masses around 10~GeV. Our full treatment of the LUX efficiencies provide more stringent limits than those from the more conservative analysis performed in that reference.} We present for the first time the LUX SD proton-only limits and discuss the general SD mixed coupling case, finding that LUX alone fully excludes the SD-interacting DM interpretation of the anomalous signal seen in DAMA \cite{Bernabei:2010mq}.

The paper is structured as follows. In \refsec{Theory} we provide a brief and self-contained review on the physics of WIMP-nuclear scattering, in both the SI and SD case. In \refsec{LUX} we review the LUX experiment, and present the methodology for our determination of the LUX likelihood and constraints. \Refsec{Results} presents our SI and SD coupling limits as a function of the WIMP mass, including comparisons with the published LUX limits (where relevant), and those of other DM experiments. We also place constraints on an effective DM theory for which the SD constraints are particularly important: the case of fermionic DM interacting with a spin-1 mediator that has an axial-vector coupling to both the WIMP and to SM fermions. Finally, we present conclusions in \refsec{Conclusions}. \Refapp{LUXCalc} describes the use of the \LUXCalc{} software.

\section{Theory review}
\label{sec:Theory}


Direct detection experiments aim to observe the recoil of a nucleus in
a collision with a DM particle~\cite{Goodman:1984dc}.  After
an elastic collision with a WIMP $\chi$ of mass $\mchi$, a nucleus of
mass $M$ recoils with energy $\Enr = (\mu^2 v^2/M)(1-\cos\theta)$,
where $\mu \equiv \mchi M/ (\mchi + M)$ is the reduced mass of the
WIMP-nucleus system, $v$ is the speed of the WIMP relative to the
nucleus, and $\theta$ is the scattering angle in the center of mass
frame.  The differential recoil rate per unit detector mass is
\begin{equation}\label{eqn:dRdEnr}
  \dRdEnr
    = \frac{\nchi}{M} \, \Big\langle v \frac{d\sigma}{d\Enr} \Big\rangle
    = \frac{2\rhochi}{\mchi}
      \int d^3v \, v f(\bv,t) \, \frac{d\sigma}{dq^2}(q^2,v) \, ,
\end{equation}
where $\nchi = \rhochi/\mchi$ is the number density of WIMPs, with
$\rhochi$ the local DM mass density; $f(\bv,t)$ is the
time-dependent WIMP velocity distribution; and
$\frac{d\sigma}{dq^2}(q^2,v)$ is the velocity-dependent differential
cross-section, with $q^2 = 2 M \Enr$ the momentum exchange in the
scatter.
In the typical case that the target material contains more than one
isotope, the differential rate is given by a mass-fraction weighted
sum over contributions from the isotopes, each of the form given by
\refeqn{dRdEnr}.
Using the form of the differential cross-section for the most
commonly assumed couplings,
\begin{equation}\label{eqn:dRdEnr2}
  \dRdEnr
    = \frac{1}{2 \mchi \mu^2} \; \sigma(q) \; \rhochi \eta(\vmin(\Enr),t) \, ,
\end{equation}
where $\sigma(q)$ is an effective scattering cross-section and
\begin{equation} \label{eqn:eta}  
  \eta(\vmin,t) \equiv \int_{v > \vmin} d^3v \, \frac{f(\bv,t)}{v}
\end{equation}
is the mean inverse speed, with
\begin{equation} \label{eqn:vmin} 
  \vmin =\sqrt{\frac{M \Enr}{2\mu^2}}
\end{equation}
the minimum WIMP velocity that can result in a recoil energy $\Enr$.
\Refeqn{dRdEnr2} conveniently factorizes the differential rate into
particle physics terms ($\sigma$) and astrophysics terms ($\rhochi\eta$),
which we describe separately in the following sections.
More comprehensive reviews of direct detection can be found in
Refs.~\cite{Lewin:1995rx,Jungman:1995df,Bertone:2004pz,Freese:2012xd}.

\subsection{Particle physics: cross-section}
\label{sec:CrossSection}

For a SUSY neutralino and many other WIMP candidates, the dominant
WIMP-quark couplings in direct detection experiments are the scalar
and axial-vector couplings, which respectively give rise to SI and SD
cross-sections~\cite{Jungman:1995df}.  In both cases,
\begin{equation}\label{eqn:dsigmadq}
  \frac{d\sigma}{dq^2}(q^2,v) = \frac{\sigma_{0}}{4 \mu^2 v^2}
                              F^2(q) \, \Theta(\qmax-q)
\end{equation}
to leading order. Here, $\Theta$ is the Heaviside step function,
$\qmax = 2 \mu v$ is the maximum momentum transfer in a collision at a
relative velocity $v$, and the requirement $q < \qmax$ gives rise to the
lower limit $v>\vmin$ in the integral for $\eta$ in \refeqn{eta}. In the
above equation, $\sigma_0$ is the scattering cross-section in the
zero-momentum-transfer limit (we shall use $\sigmaSI$ and $\sigmaSD$
to represent this term in the SI and SD cases, respectively) and
$F^2(q)$ is a form factor to account for the finite size of the
nucleus.  The WIMP coherently scatters off the entire nucleus when the
momentum transfer is small, giving $F^2(q) \to 1$.  However, as the
de~Broglie wavelength of the momentum transfer becomes comparable to
the size of the nucleus, the WIMP becomes sensitive to the spatial
structure of the nucleus and $F^2(q) < 1$, with $F^2(q) \ll 1$ at
higher momentum transfers. It is traditional to define a form-factor
corrected cross-section
\begin{equation}\label{eqn:sigmaq}
  \sigma(q) \equiv \sigma_0 F^2(q) \, ,
\end{equation}
as was used in \refeqn{dsigmadq} above.  We note that this is an
\textit{effective} cross-section, whereas the \textit{actual}
scattering cross-section is given by $\int dq^2
\frac{d\sigma}{dq^2}(q^2,v)$ for a given relative velocity $v$.
The total WIMP-nucleus scattering rate is then the sum over both the
SI and SD contributions, each with its own form factor.

\subsubsection{Spin-independent cross-section (SI)}
\label{sec:SI}

The SI WIMP-nucleus interaction, which occurs through operators such
as $(\bar{\chi}\chi)(\bar{q}q)$, has the cross-section 
\begin{equation} \label{eqn:sigmaSI}
  \sigmaSI
     \ =\ \frac{\mu^2}{\pi} \, \Big[ Z \GpSI + (A-Z) \GnSI \Big]^{2}
     \ =\ \frac{4\mu^2}{\pi} \, \Big[ Z \fpSI + (A-Z) \fnSI \Big]^{2} \; ,
\end{equation}
where $Z$ and $A-Z$ are the number of protons and neutrons in the
nucleus, respectively, and $\fpSI$ ($\fnSI$) is the effective coupling
to the proton (neutron), with the alternate normalization $\GNSI=2\fNSI$
also found in the literature.\footnote{%
  Notably, $\GNSI$ are the $G_F$-like effective four-fermion coupling
  constants in the case of scalar interactions \cite{Gondolo:1996qw}
  and are the normalization used in \DarkSUSY{} \cite{Gondolo:2004sc}.
  \MicrOMEGAs{} uses $\lambda_{N} = \frac{1}{2}\,\GNSI$ \cite{Belanger:2008sj}.
  }
For neutralinos and most other WIMP
candidates with a SI interaction arising through scalar couplings,
$\fnSI \simeq \fpSI$ and the SI scattering cross-section of WIMPs with
protons and neutrons are roughly comparable, $\sigmanSI \approx
\sigmapSI$.  For identical couplings ($\fnSI = \fpSI$), the SI
cross-section can be written as
\begin{equation} \label{eqn:sigmaSI2}
  \sigmaSI = \frac{\mu^2}{\mup^2} A^2 \, \sigmapSI \, ,
\end{equation}
where $\mup$ is the WIMP-proton reduced mass.  As neutralinos are the
currently favored WIMP candidate, this assumption is widely made
throughout the direct detection literature, though models can be
constructed that violate this $\fnSI \simeq \fpSI$ condition
(\eg\ isospin-violating DM \cite{Feng:2011vu}). We will
consider only identical SI couplings case when later examining the
LUX results.

The SI cross-section grows rapidly with nuclear mass due to the $A^2$
factor in \refeqn{sigmaSI2}, which arises from the fact that the total
SI coupling of the WIMP to a nucleus is a coherent sum over the
contributions from individual protons and neutrons within. Direct
detection experiments therefore often use heavy nuclei to increase
their sensitivity to WIMP scattering.

The SI form factor is essentially a Fourier transform of the mass
distribution of the nucleus.  A reasonably accurate approximation is
the Helm form factor \cite{Helm:1956zz,Lewin:1995rx}:
\begin{equation} \label{eqn:SIFF}
  F(q) = 3 e^{-q^2 s^2/2} \; \frac{\sin(qr_n)- qr_n\cos(qr_n)}{(qr_n)^3} \, ,
\end{equation}
where $s\simeq 0.9$~fm and $r_n^2 = c^2 + \frac{7}{3} \pi^2 a^2 - 5
s^2$ is an effective nuclear radius with $a \simeq 0.52$~fm and $c
\simeq 1.23 A^{1/3} - 0.60$~fm.  Further details on SI form factors
can be found in Refs.~\cite{Lewin:1995rx,Duda:2006uk}.

\subsubsection{Spin-dependent cross-section (SD)}
\label{sec:SD}

SD scattering is due to the interaction of a WIMP with the spin of the
nucleus through operators such as $(\bar{\chi}\gamma_{\mu}\gamma_5
\chi)(\bar{q} \gamma^{\mu} \gamma_5 q)$, and takes place only in those
detector isotopes with an unpaired proton and/or unpaired neutron.
The SD WIMP-nucleus cross-section is
\begin{align} \label{eqn:sigmaSD}
  \begin{split}
    \sigmaSD
      \ &=\ \frac{4 \mu^2}{\pi} \, \frac{(J+1)}{J} \,
            \Big[ \GpSD \Sp + \GnSD \Sn \Big]^2 \\
      \ &=\ \frac{32 \mu^2 \, G_F^2}{\pi} \, \frac{(J+1)}{J} \,
            \Big[ \apSD \Sp + \anSD \Sn \Big]^2 \, ,
  \end{split}
\end{align}
where $G_F$ is the Fermi constant, $J$ is the spin of the nucleus,
$\Sp$ and $\Sn$ are the average spin contributions from the
proton and neutron groups, respectively, and $\apSD$ ($\anSD$) are the
effective couplings to the proton (neutron) in units of $2\sqrt{2}G_F$;
the alternative normalization $\GNSD = 2\sqrt{2}G_F\,\aNSD$ is also
found in the literature.\footnote{%
  Though we use $\aNSD$ and $\GNSD$ to distinguish between the two
  normalizations here, $\aNSD$ is frequently used within the literature
  for \textit{both} cases. The $\GNSD$ normalization is used by
  \DarkSUSY{} \cite{Gondolo:1996qw,Gondolo:2004sc}, while \micrOMEGAs{}
  uses $\xi_{N} = \frac{1}{2}\,\GNSD$ \cite{Belanger:2008sj}.
  }
Unlike the SI case, the
two SD couplings $\apSD$ and $\anSD$ may differ substantially (though
they are often of similar order of magnitude), so that a
simplification comparable to \refeqn{sigmaSI2} is
not made in the SD case. Because of the uncertain theoretical
relation between the two couplings and following from the fact that
one of $\Sp$ or $\Sn$ is often much smaller than the other,
experiments typically only significantly constrain one of the two SD
cross-sections, $\sigmapSD$ or $\sigmanSD$, but not both.

The SD form factor is given in terms of the structure function $S(q)$ as
\begin{equation}
  F^2(q)=\frac{S(q)}{S(0)}
\end{equation}
such that $F^2(0)=1$ as previously prescribed. 
The $S(q)$ have the functional form
\begin{equation}\label{eqn:Sq}
  S(q) = \apSD^2\,\Spp(q) + \anSD^2\,\Snn(q) + \apSD\;\!\anSD\;\!\Spn(q) \;.
\end{equation}
The differential SD scattering cross-section can alternatively be written
in terms of this structure function as\footnote{%
  The factor of $8G_F^2$ is omitted if \refeqn{Sq} is written in terms
  of $\GNSD$ rather than $\aNSD$.  Again, sometimes the latter
  notation is used in the literature to represent the former
  normalization as defined here.}%
\begin{equation}\label{eqn:dsigmadqS}
  \frac{d\sigma}{dq^2}(q^2,v) = \frac{8 G_F^2}{v^2\,(2J+1)} \, S(q) \, \Theta(\qmax-q) \;.
\end{equation}
In the limit of zero momentum transfer ($q\rightarrow 0$) the functions $S_{xy}(0)\;(x,y=\mathrm{p,n})$ can be related to expectation values of the proton/neutron spin  \cite{Bednyakov:2004xq,Bednyakov:2006ux}:
\begin{equation}
  \begin{split}
    \Sp^2 &= \frac{J}{(J+1)}\frac{\pi}{(2J+1)} \, \Spp(0) \quad\text{and} \\
    \Sn^2 &= \frac{J}{(J+1)}\frac{\pi}{(2J+1)} \, \Snn(0)\;.
  \end{split}
\end{equation}
Under an alternative basis
\begin{equation}
  a_0=\apSD+\anSD \qquad a_1=\apSD-\anSD \;,
\end{equation}
which is a more convenient basis for nuclear physics work,
\begin{equation}
  S(q)=a_0^2\,S_{00}(q)+a_1^2\,S_{11}(q)+a_0\;\!a_1\;\!S_{01}(q)\;,
\end{equation}
where the structure functions in the two bases are related by
\begin{align}
  \begin{split}
    \Spp &= S_{00}+S_{11}+S_{01}  \\
    \Snn &= S_{00}+S_{11}-S_{01}  \\
    \Spn &= 2\left(S_{00}-S_{11}\right)\;.
  \end{split}
\end{align}
The $S_{ij}\;(i,j=0,1)$ are calculated in the literature. For the two non-zero-spin xenon isotopes ($^{129}$Xe and $^{131}$Xe), we take these structure functions from Ref.~\cite{Menendez:2012tm}, utilizing one plus two body (1b+2b) axial-vector currents where possible.

\subsection{Astrophysics: dark matter distribution}
\label{sec:VelocityDist}

The DM halo in the local neighborhood is most likely
dominated by a smooth and well-mixed (virialized) component with an
average density $\rhochi$.
The simplest model for this smooth component is often taken to be the
Standard Halo Model (SHM)~\cite{Drukier:1986tm,Freese:1987wu}, a non-rotating
isothermal sphere with an isotropic, Maxwellian velocity distribution
and most probable speed $v_0$, where for the SHM, $v_0$ is equal to the disk rotation speed $\vrot$.  The SHM velocity distribution in the Galactic (\ie\ non-rotating) rest frame is
\begin{equation} \label{eqn:TruncMaxwellian}
  \widetilde{f}(\bv) =
    \begin{cases}
      \frac{1}{\Nesc} \left( \frac{1}{ \pi v_0^2} \right)^{3/2}
        \, e^{-\bv^2\!/v_0^2} , 
        & \textrm{for} \,\, |\bv| < \vesc  \\
      0 , & \textrm{otherwise}.
    \end{cases}
\end{equation}
Here,
\begin{equation} \label{eqn:Nesc}
  \Nesc = \erf(z) - \frac{2}{\sqrt{\pi}} z e^{-z^2} \, ,
\end{equation}
with $z \equiv \vesc/\vmp$, is a normalization factor. The Maxwellian distribution is truncated
at the escape velocity $\vesc$ to account for the fact that WIMPs with
sufficiently high velocities escape the Galaxy's potential well.  

Ultimately, the velocity distribution in the Earth's rest frame is the one relevant for direct detection, which can be obtained after a Galilean boost:
\begin{equation} \label{eqn:vdist}
  f(\bv) = \widetilde{f}(\bvsun + \bv) \, ,
\end{equation}
where $\bvsun = \bvLSR + \bvsunpec$ is the motion of the Sun relative to the Galactic rest frame, $\bvLSR = (0,\vrot,0)$ is the motion of the Local Standard of Rest in Galactic coordinates,\footnote{Galactic coordinates are aligned such that $\hat{\mathbf{x}}$ is the direction to the Galactic center, $\hat{\mathbf{y}}$ is the direction of the local disk rotation, and $\hat{\mathbf{z}}$ is orthogonal to the plane of the disk.} and $\bvsunpec$ is the Sun's peculiar velocity. The additional time-dependent orbital motion of the Earth about the Sun is postulated to give rise to a measurable modulation in the signal \cite{Freese:1987wu,Freese:2012xd}; indeed, some experiments have claimed positive results for such signatures \cite{Bernabei:2000qi}. This small Earth orbital motion and resulting time dependence has been neglected here as it is not relevant to the LUX calculations. 

\begin{table*}
  \begin{center}
  \addtolength{\tabcolsep}{0.5em}
  \begin{tabular}{lcccc}
    \hline 
    & estimate & canonical & \LUXCalc{} default & Refs. \\
    $\rhochi$ [GeV/cm$^3$] & 0.2 -- 0.7 & 0.3 & 0.4 &
      \cite{Caldwell:1981rj,Catena:2009mf,Weber:2009pt,Salucci:2010qr,Pato:2010yq,Bovy:2012tw} \\
    $\vrot$ [km/s]         & 200 -- 250 & 220 & 235 &
      \cite{Kerr:1986hz,Reid:2009nj,McMillan:2009yr,Bovy:2009dr} \\
    $\vesc$ [km/s]         & 500 -- 600 & 544 & 550 &
      \cite{Smith:2006ym} \\
    \hline 
  \end{tabular}
  \end{center}
  \caption[SHM parameters]{%
    Values of parameters describing the Standard Halo Model (SHM).
    The columns show in order:
    ranges of values typically found in the literature,
    commonly used values for comparison of experimental results,
    and the default values used by \LUXCalc{}.
    In the SHM, the final parameter $v_0$ is not independent, but
    fixed by the relation $v_0 = \vrot$.
    }
  \label{tab:SHM}
\end{table*}

The choice of values for the various DM distribution parameters is important for interpreting direct detection results. Due to the inability to directly observe the DM and various systematic issues in interpreting observations of the Galaxy, some of those parameters are known to limited precision. The range of estimates for these parameters are shown in \reftab{SHM}, as well as values commonly used for comparing direct detection results (``canonical'') and the default values used by \LUXCalc{}. The \LUXCalc{} default values for $\rhochi$ and $\vrot$ are somewhat larger than the historical canonical values as more recent estimates tend to prefer the somewhat larger values. However, the canonical values are not inconsistent with the current observations and will thus continue to see wide usage. The parameter values can be easily changed in \LUXCalc{} and we will use the canonical values ourselves when comparing LUX results with other experiments in later sections. In addition to these SHM parameters, we take $\bvsunpec = (11,12,7)$~km/s \cite{Schoenrich:2009bx}.

\section{The LUX experiment and analysis}
\label{sec:LUX}

In this section, we describe the basic operation of the LUX DM
search detector and how their data is used to constrain WIMP parameters.
The LUX experiment uses a liquid-xenon time projection chamber (TPC) to
identify DM recoil events and distinguish them from other
(background) events \cite{Akerib:2012ys}.
The first LUX results, released in 2013 \cite{Akerib:2013tjd}, involved
a fiducial exposure of $1.01\times10^{4}$~kg-days, comparable to that
of the then-leading\footnote{%
  In terms of sensitivity to WIMPs with SI interactions
  and masses above $\sim$10~GeV.
  }
XENON100 experiment \cite{Aprile:2012nq} ($7.6\times10^{3}$~kg-days),
which likewise used a TPC detector with a xenon target.
The slightly larger exposure and somewhat better detector performance
characteristics allowed LUX to overtake XENON100 and LUX is now
the leading experiment in terms of sensitivity to a large variety of
WIMPs (by about a factor of two over XENON100).

The principles of operation for a TPC are as follows.  A recoiling xenon
nucleus (or recoiling electron, in the case of some background
processes) in the liquid xenon target will interact with other nearby
atoms, inducing both excitations and ionization of those atoms.
The relatively quick relaxation of the excitations releases photons
that are collected by photomultiplier tubes (PMTs);
this prompt scintillation is referred to as the S1 signal.
While some of the ionized electrons can recombine (possibly producing
excitations and contributing to the S1 prompt scintillation signal),
many of the free electrons are drawn away from the interaction site to
the surface of the liquid by the application of an electric field
across the liquid.  Above the liquid is a small region of gaseous xenon
under a higher electric field.  Electrons reaching this region rapidly
accelerate and collide with xenon atoms in the gas, releasing photons
and more electrons in a cascade process; this ``proportional
scintillation'' is the S2 signal.  The drift time of the ionized
electrons across the liquid is substantially longer than the relaxation
time of the xenon excitations, so the S1 and S2 scintillation signals
are easily distinguishable.  The benefit of observing two signals is
that electron recoils, produced by background radiation, tend to
produce a relatively larger amount of S2 for a given S1, so S2/S1 is
used as a background discrimination parameter.

For a given WIMP spectrum, the average expected number of signal events
in some analysis region is
\begin{equation} \label{eqn:signal}
  \mu = MT \int_0^{\infty} dE \; \phi(E) \, \dRdE(E) \, ,
\end{equation}
where $MT$ is the detector mass$\times$time exposure and $\phi(E)$ is
the fraction of recoil events of energy $E$ that will be both observed
and fall into the predefined analysis region.  This $\phi(E)$ detection
efficiency accounts for various trigger efficiencies, intrinsic
statistical fluctuations and the PMT response (\ie\ detector resolution),
and analysis cuts.
The benefit of the above form is that all the complicated detector
physics and responses are rolled into $\phi(E)$, which is independent of
the type of WIMP interaction or spectrum.  Thus, for a given experimental
result and analysis region, $\phi(E)$ can be tabulated once and then
used for analyzing arbitrary WIMPs.
We use the \TPCMC{} monte carlo \cite{Savage:2015tpcmc} to model the
detector response and generate the relevant $\phi(E)$ efficiency
curves.
\TPCMC{} relies on \NEST{} \cite{Szydagis:2011tk,Szydagis:2013sih,NEST:url}
for modeling the microphysics of a recoiling xenon atom.

We take as our analysis region in the S2/S1-S1 plane the
region above the S2 $\ge$ 200~PE threshold, below the nuclear recoil
calibration data mean S2/S1 curve, and 2~PE~$\le$ S1 $\le$~30~PE.
This region matches that used by the LUX collaboration, except for the
imposition of the hard S2/S1 cut that is not necessary for their
profile likelihood analysis \cite{Cowan:2010js}.  This region contains
one observed event at an S1 of 3.1~PE with a mean expected background
of $b = 0.64$~events.  Ideally, a lower S2/S1 bound to the region
should be imposed in the count-based analyses we will be using, but
such a bound can often be set low enough that, in practice, the $\phi(E)$
are negligibly affected.  The lower bound would serve more to exclude
very low S2/S1 events that are almost certainly backgrounds; luckily,
there are no such events in the LUX results and this issue is
moot.\footnote{%
  A conservative imposition of a lower S2/S1 cut corresponding to the
  10\% lower tail of the nuclear recoil calibration data (dashed red
  curve in Figure~4 of Ref.~\cite{Akerib:2013tjd}) results in constraints
  that are weaker by $\sim$ 20-30\%.
  }

Given an observed number of events $N$ and expected background $b$ --- being
1 and 0.64 for this LUX analysis, respectively --- a likelihood can be
constructed from the Poisson distribution, with
\begin{equation} \label{eqn:Poisson}
  \mathcal{L}(\mchi,\sigmas|N)
    = P(N|\mchi,\sigmas)
    = \frac{(b+\mu)^N \, e^{-(b+\mu)}}{N!} \, ,
\end{equation}
where $\mu = \mu(\mchi,\sigmas)$ is the expected number of signal events
for a given WIMP mass $\mchi$ and one or more scattering cross-sections
$\sigmas$ (or, alternatively, couplings).  This likelihood can be
easily combined with those from a variety of other physics data, such
as from colliders or indirect DM searches, in statistical scans of the
Minimal Supersymmetric Standard Model (MSSM) or other new physics
frameworks; see \eg\ \cite{Akrami:2010dn,Balazs:2013qva,Strege:2012bt,
Buchmueller:2013rsa,Kowalska:2014hza}.

We derive constraints in a $\sigma$-$\mchi$ parameter space via two
methods: the Feldman-Cousins (Poisson-based) method \cite{Feldman:1997qc}
and Yellin's maximum gap method \cite{Yellin:2002xd}.  These two
techniques correspond to analyses with and without background
subtraction, respectively, with the latter case commonly found in the
literature due to the difficulty in reliably identifying and
characterizing background contributions.

The Feldman-Cousins method derives a confidence interval in $\sigma$ for
each $\mchi$ (resulting in a raster scan in the $\sigma$-$\mchi$ plane)
that is consistent with the observed number of events given the expected
background.  This confidence interval may be either one or two sided
and, thus, is capable of excluding the zero-signal case when excess
events are found.  This method is relatively straightforward and easy
to implement, but one of the drawbacks is the lack of spectral
information in the analysis: only the total counts are used, not the
distribution of S1 values (a coarse proxy for recoil energy).  This
spectral information can be useful in distinguishing between signal
and background and can aid in constraining the mass of the WIMP in the
event of a positive result, as heavier WIMPs induce more energetic
xenon recoils and larger S1 values, on average.  Analyses that make use
of spectral information are substantially more complex to implement and
are thus typically only performed by the experimental collaborations
themselves; see \eg\ Refs.~\cite{Aprile:2011hx,Aprile:2012vw}.

The maximum gap method makes no presumptions about the amount of
background that might be contributing to the observed events, instead
assuming any or all of the events might be signal to generate a
conservative exclusion limit in $\sigma$ at each $\mchi$:  any
cross-sections above this limit would yield too many events even if
background contributions were ignored.  This method does make use of
the S1 distribution of the observed events, focusing on an S1 range
where the expected number of events is largest relative to the number
of observed events.  Specifically, the maximum gap method breaks the
full observable (S1) range into intervals separated by the observed
events.  If $\mu_k$ are the predicted number of events in each of these
intervals, the ``maximum gap'' is the one where $\mu_k$ is largest.
If $x \equiv \frac{\mu_{k,\max}}{\mu}$ is the fraction of signal events
expected in this interval, then
\begin{equation} \label{eqn:MaxGapC0}
  C_0(x,\mu) = \sum_{k=0}^{\lfloor \mu/x\rfloor} \frac{(kx-\mu)^k\,e^{-kx}}{k!}
               \left( 1 + \frac{k}{\mu-kx} \right)
\end{equation}
is the probability of the maximum gap being smaller than observed.
In other words, a WIMP mass and cross-section(s) is excluded at greater
than a 90\% confidence level (CL) if $C_0 \ge 0.9$.  To perform this
calculation with the LUX result (with one event at an S1 of 3.1~PE), we
divide the previously discussed LUX S2/S1-S1 analysis region into two
parts, S1$\,\in\,$[2,3.1]~PE and S1$\,\in\,$[3.1,30]~PE, and use \TPCMC{} to
generate efficiency factors $\phi_1(E)$ and $\phi_2(E)$, respectively,
for the two intervals.\footnote{%
  By definition, $\phi(E) = \sum_k \phi_k(E)$.%
  }
The expected number of events in an interval $\mu_k$ can then be
calculated via \refeqn{signal} under the replacement $\phi\to\phi_k$.

In tandem with this paper, we provide \LUXCalc, a software package for
performing the above-described LUX analyses.  \LUXCalc{} can be used as
a standalone program or as a library to be called from other software
packages such as \DarkSUSY{} \cite{Gondolo:2004sc,DarkSUSY:url} and
\micrOMEGAs{} \cite{Belanger:2001fz,Belanger:2013oya,MicrOMEGAs:url}.
A description of \LUXCalc{} and its usage can be found in \refapp{LUXCalc}.
The software can be found at Ref.~\cite{LUXCalc:url} or as ancillary
files to the arXiv version of this paper.

Before turning to our results, we take a moment to stress that the
$\phi(E)$ and $\phi_k(E)$ curves used here require a full statistical
modeling of LUX's TPC detector to generate and cannot be trivially
generated from any efficiencies provided by LUX in
Ref.~\cite{Akerib:2013tjd}.  One might be tempted to take the no-S2/S1
cut efficiency from Figure~1 of that reference and apply an additional
factor of 0.5 to account for the fraction of nuclear recoils falling
below the S2/S1 mean in the calibration data.  There are two reasons
why this is incorrect.  First, the S2/S1 cut is not independent of the
other cuts and, in fact, is very highly correlated with the S1 $\in$
[2,30]~PE cut near the boundaries.  Second, the mean of the
calibration recoil band in S2/S1-S1 space represents the distribution
convolved over energy for that particular calibration spectrum; this
does not imply that 50\% of the events at any specific energy will
fall below that mean.  Finally, the efficiency curve provided by LUX
(without the S2/S1 cut), applies only to the whole [2,30]~PE S1
interval and cannot be decomposed into the subintervals used for the
maximum gap analysis.  As we show in the next section, our efficiency
curves nearly exactly reproduce LUX's own low-mass constraints, an
indication that those efficiency curves are correctly modeled (the
naive approach of simply applying an additional 50\%
nuclear-recoil-median-cut acceptance would lead to constraints too
weak by a factor of two at low masses).
A tabulation of these efficiencies is also included as an ancillary
file on the arXiv version of this paper.

\section{Physics results}
\label{sec:Results}

In this section, we apply the previously described methods to analyze
various physics models.
We first examine in \refsec{GenericResults} the LUX constraints on
generic WIMPs with SI and SD interactions,
comparing our results with those from other DM searches and
from the LUX collaboration itself.  In \refsec{EffectiveTheory}, we then
apply LUX constraints to an effective theory model where SD
interactions are particularly relevant.

\subsection{Generic coupling limits}
\label{sec:GenericResults}

\begin{figure}
  \insertfig{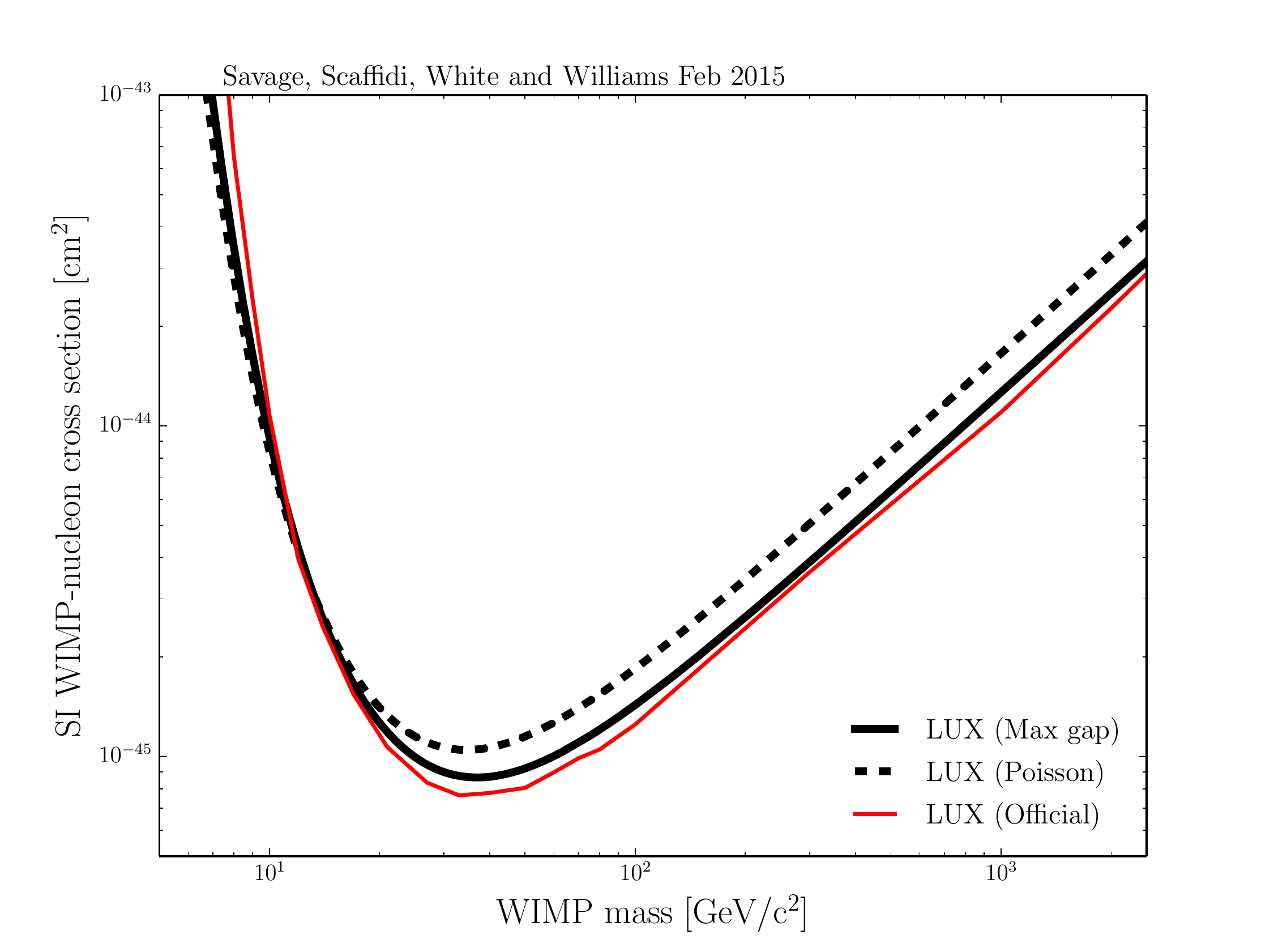}
  \caption{
    Spin-independent cross-section constraints for the LUX experiment.
    Constraints are generated with and without background subtraction:
    a Poisson-based analysis with one observed event and 0.64 expected
    background events is used in the former case (dashed black), while
    the maximum gap method is used in the latter case (solid black);
    see the text for details.  For comparison, the official LUX
    collaboration constraints are also shown (red), based upon an
    event-likelihood analysis.
    All constraints are at the 90\%~CL; cross-sections above the curves
    are excluded at greater than this level.
    }
  \label{fig:SI}
\end{figure}

\begin{figure}
  \insertfig{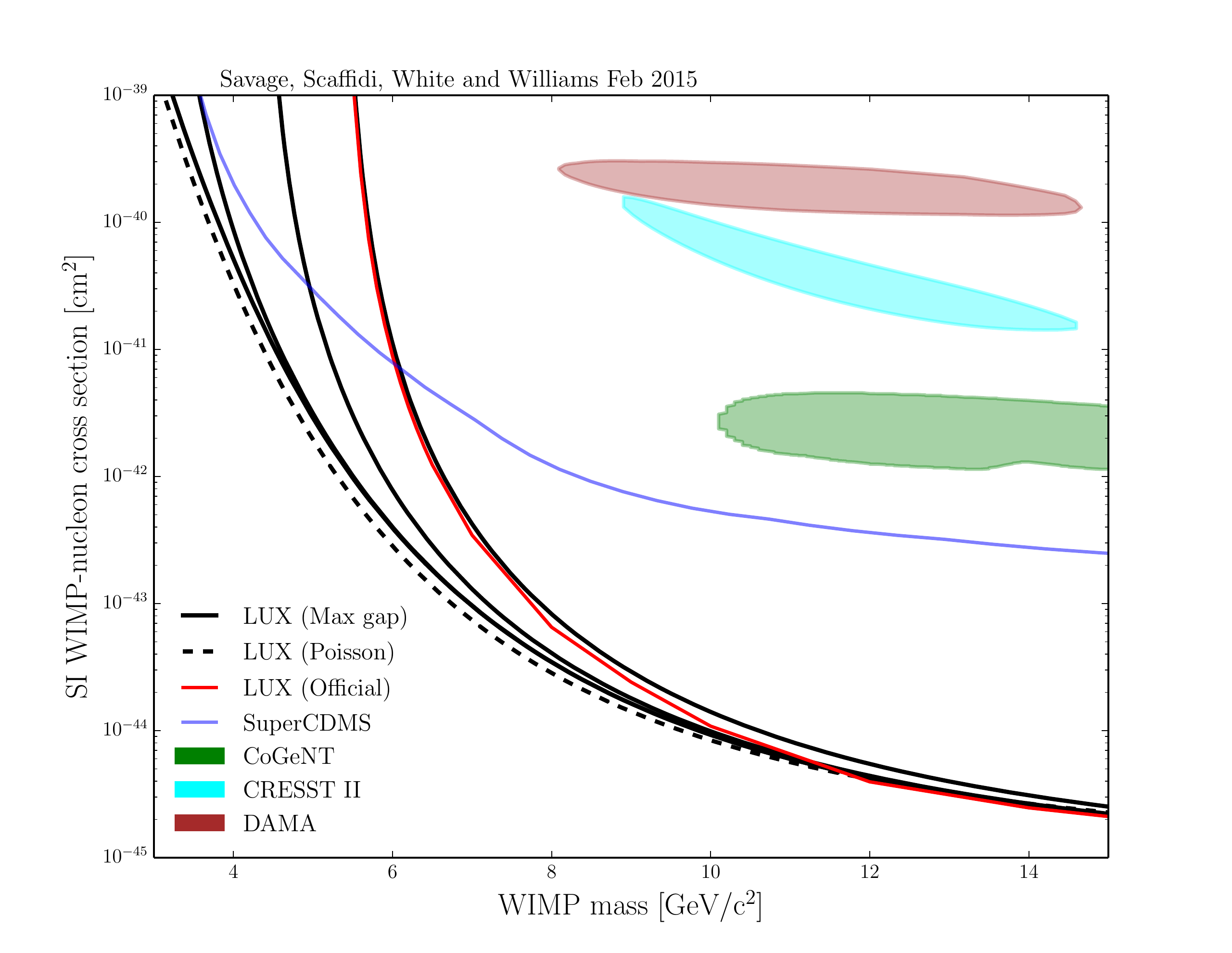}
  \caption{
    Spin-independent cross-section constraints from various direct
    searches.
    Parameters that can reproduce the CoGeNT (90\%~CL), CRESST (2$\sigma$), and DAMA (3$\sigma$)
    anomalous signals are shown in the filled regions.
    SuperCDMS and LUX exclusion constraints are shown at the 90\%~CL.
    The LUX maximum gap constraints (solid black) are shown with
    nuclear recoil spectra limited to $E > 0,1,2,3$~keV (strongest to
    weakest).  The official LUX limit used a conservative 3~keV minimum.
    }
  \label{fig:SIlowmass}
\end{figure}

Our LUX SI scattering constraints are shown in \reffig{SI},
with the maximum gap limit shown in solid black and the Poisson-based
constraint shown in dashed black; constraints are at the 90\%~CL.
The official LUX limits are also
shown (thin red).  Though they are not our preferred parameters, we use
$\vmp = \vrot = 220$~km/s and $\rhochi = 0.3$~GeV/cm$^3$ to generate
constraints in this and later figures to allow for a proper
comparison with various other experimental constraints that use these
values, such as the official LUX limits.  The maximum gap limit is
remarkably close to the LUX collaboration's own limit, while the
likelihood-based constraint is weaker by $\sim$30\% except at lower
WIMP masses (which will be discussed momentarily).\footnote{%
  While the likelihood analysis can produced two-sided limits, in this
  case the lower limit is simply zero cross-section, \ie\ no WIMP
  signal: the LUX result is consistent with backgrounds alone at the
  90\%~CL in this analysis.
  }
There are two main potential reasons why the likelihood-based constraint
is weaker: (1) the analysis region for our analysis is the lower-half
of the nuclear recoil band and thus contains only about half of the
potential signal (while the LUX collaboration analysis uses all of it)
and (2) our Poisson likelihood makes no use of spectral information
(\ie\ event S1).  The first reason is not, in fact, a major issue in
practice as the upper half of the nuclear recoil band that is being
ignored is contaminated by background events and does not significantly
improve the signal-to-noise in those analyses that use it.

The LUX SI scattering constraints for low WIMP masses are shown in
\reffig{SIlowmass}.  For comparison, the parameters consistent with a
DM interpretation of the anomalous signals seen in
CoGeNT \cite{Aalseth:2014jpa}, CRESST \cite{Angloher:2011uu}, and
DAMA \cite{Bernabei:2010mq,Kelso:2013gda} are shown, as well as the
exclusion constraints from SuperCDMS \cite{Agnese:2014aze}.
As the WIMP mass is lowered, the Poisson-based likelihood constraint
(dashed black curve) becomes comparable to and then slightly better than
the maximum gap limit (farthest left solid black curve, other curves
discussed below).
The low-mass improvement of the likelihood case relative to the maximum
gap case is due to the fact that the single event in the analysis
becomes consistent with the expected WIMP spectrum, leading to a slight
weakening in the sensitivity of the maximum gap method.

As measurements of the scintillation and ionization at very low nuclear
recoil energies are limited and the theoretical models will eventually
break down at sufficiently low energies (see \eg\ 
Ref.~\cite{Sorensen:2014sla}), the LUX collaboration
conservatively ignores contributions from WIMP scattering events with
$\Enr < 3$~keV when analyzing their results.  While this has little
impact on constraints for WIMPs heavier than $\sim$20~GeV, it becomes
important for light WIMPs as light WIMPs can only induce low-energy
recoils.  In the figure, we show our own maximum gap constraints when
considering only $\Enr \ge 0$, 1, 2, and 3~keV (solid black curves
from left to right, or most to least constraining).  With the same
$\Enr \ge 3$~keV that LUX uses, the maximum gap constraint closely
matches their constraint.
To be clear, placing a minimum on the contributing $\Enr$ is not quite
the same as defining the threshold in the detector.  The actual trigger
and S1 analysis thresholds are already built into the $\phi(\Enr)$
efficiency term in \refeqn{signal} regardless of the choice of lower
bound in the integral over $\Enr$.  That efficiency falls rapidly below
3~keV, from 23\% at 3~keV, to 4\% at 2~keV and 0.04\% at 1~keV.  Even
if the lower bound of integration is set to $\Enr=0$~keV, no recoil
events with energies below 0.5~keV will contribute to the signal as
$\phi(\Enr)$ is zero at these energies.
Placing a minimum requirement on $\Enr$ serves to avoid the (already
suppressed) contributions from events where the NEST model has little
experimental data to ensure its accuracy.
For our remaining analyses, we do not apply this artificial cut on the
low-energy recoil spectrum, though this choice has little impact on
our conclusions and constraints with a cut applied can be easily
generated with the \LUXCalc{} code as described in \refapp{LUXCalc}.

\begin{figure}
  \insertfig{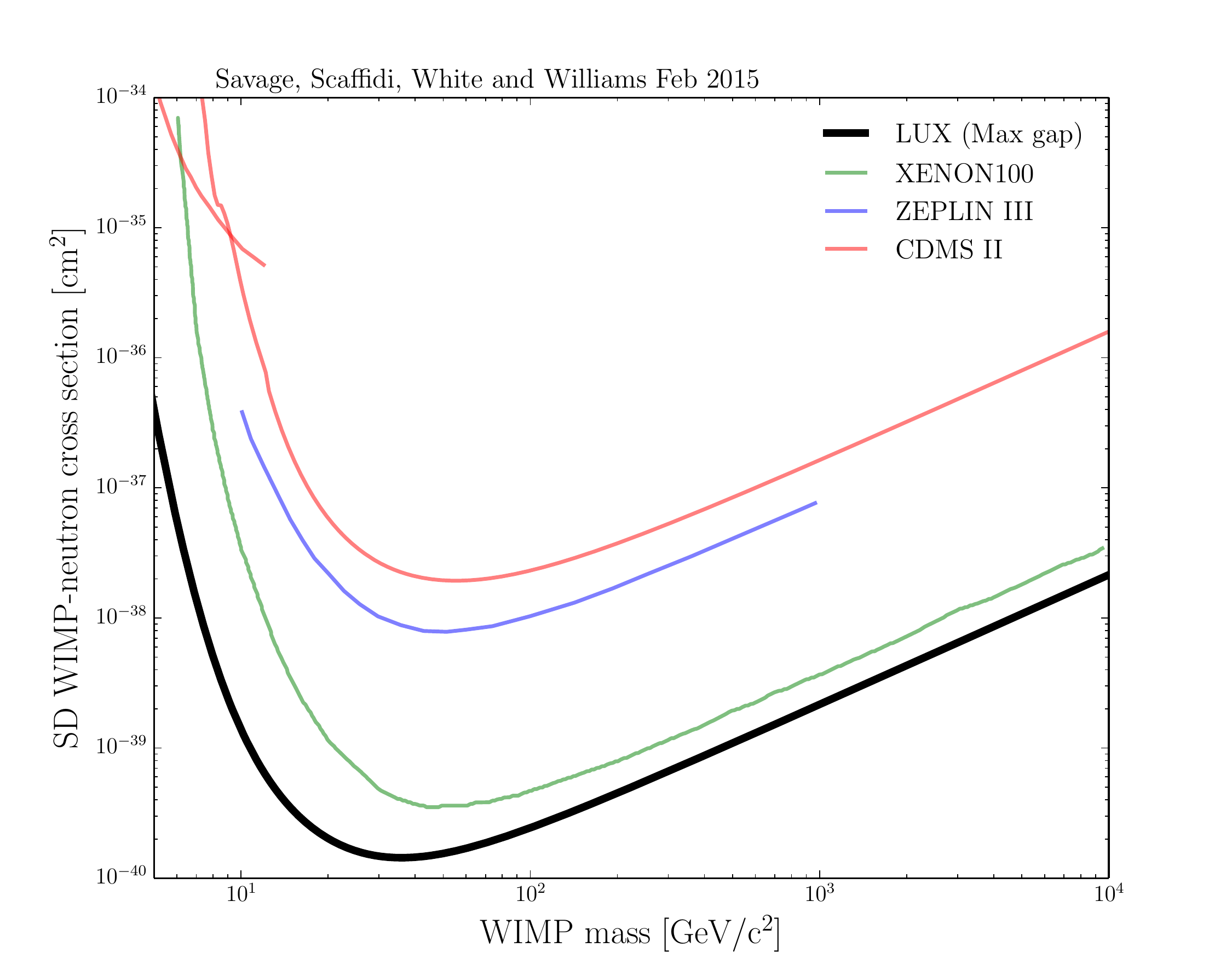}
  \caption{
    Spin-dependent WIMP-neutron cross-section constraints for the
    neutron-only coupling case.
    }
  \label{fig:SDn}
\end{figure}

\begin{figure}
  \insertfig{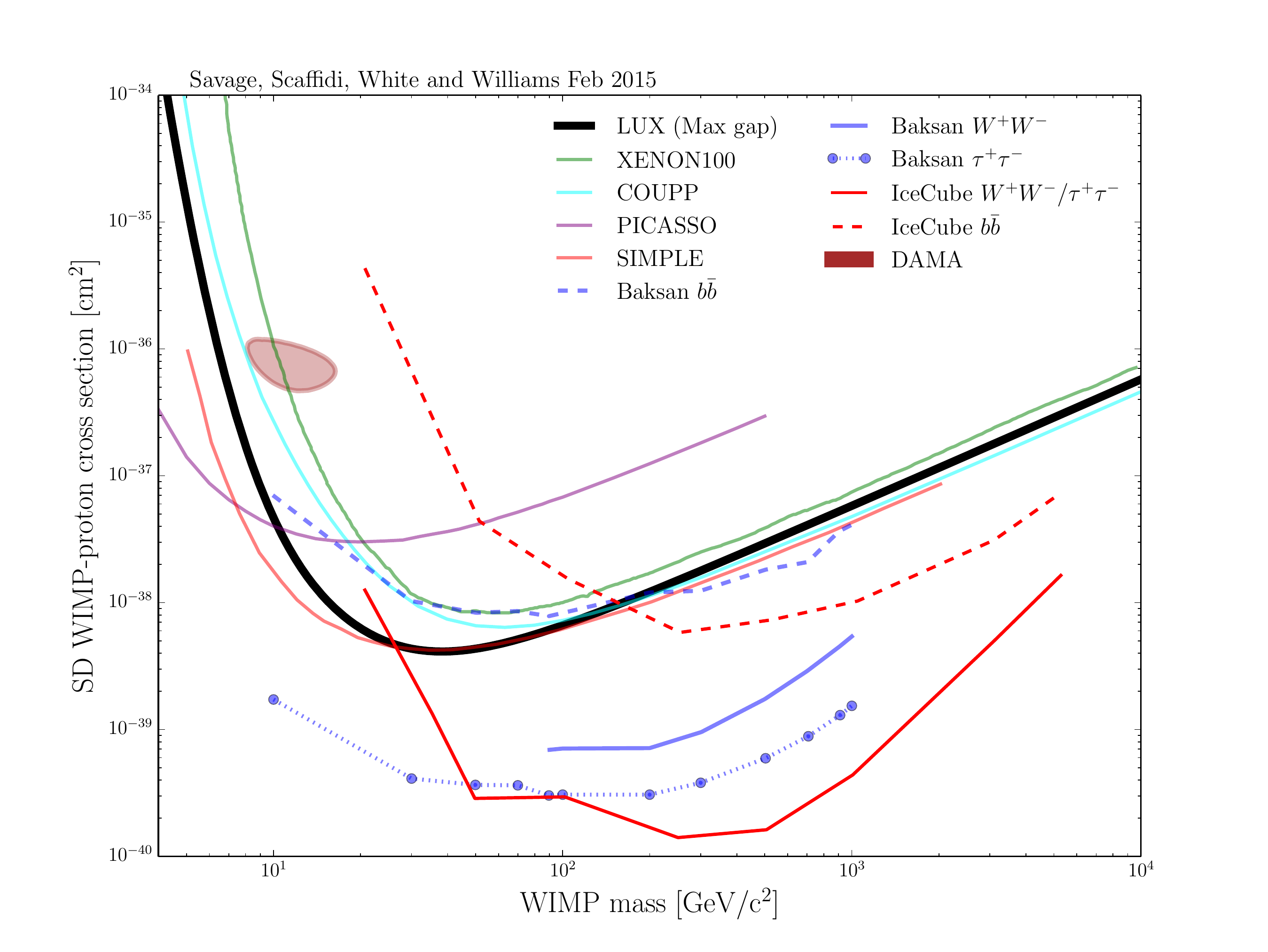}
  \caption{
    Spin-dependent WIMP-proton cross-section constraints for the
    proton-only coupling case, including indirect search limits from the
    IceCube/DeepCore experiment.
    Note the IceCube/DeepCore $W^+W^-$ constraint uses the tau channel
    for masses smaller than that of the W~boson.
    }
  \label{fig:SDp}
\end{figure}

We now turn to the SD case, with WIMP-nucleon cross-section constraints
shown in Figures~\ref{fig:SDn} \&~\ref{fig:SDp} for neutron-only
($\apSD=0$) and proton-only ($\anSD=0$) couplings, respectively.
As xenon has neutron-odd isotopes, LUX should be particularly
sensitive to a WIMP with neutron-only SD couplings. In \reffig{SDn}, we
show by the black curve the LUX constraints in this case, as determined
via the method described in \refsec{LUX}. We also show constraints from
other experiments with neutron-odd target materials:
CDMS~II \cite{Ahmed:2008eu} (at lower masses, also for the low-threshold
analysis \cite{Ahmed:2010wy}), ZEPLIN-III \cite{Akimov:2011tj}, and
XENON100 \cite{Aprile:2013doa}. Here, the somewhat improved exposure and
threshold of LUX over that of XENON100 is evident by the $\sim\times$2
stronger constraint at heavy WIMP masses and the extension of the
constraints to lower WIMP masses before losing sensitivity.

A SD proton-only interaction has historically been a means of producing
DAMA's anomalous signal while evading the null results of other
searches \cite{Savage:2008er}: the proton-even target isotopes used by
many experiments have suppressed interactions (and thus little expected
signal) in this case, while the proton-odd sodium-iodide target used
by DAMA ensures they remain sensitive to the WIMP. Recent results from
the SIMPLE \cite{Felizardo:2011uw}, PICASSO \cite{Archambault:2012pm},
and COUPP \cite{Behnke:2012ys} experiments, which also have proton-odd
target isotopes, are now in conflict with a SD proton-only coupling
explanation for the DAMA signal \cite{Bernabei:2010mq,Kelso:2013gda} as
shown in \reffig{SDp}. Due to the ever-increasing detector sizes, even
experiments with proton-even targets like XENON100 \cite{Aprile:2013doa}
are starting to probe DAMA's preferred parameter region. Our determination
of the LUX constraints are shown by the thick black curve. We see here
that LUX, even though it uses a proton-even xenon target material,
fully excludes the DAMA region.  

Indirect DM searches via neutrinos produced when WIMPs are
caught and then annihilate in the Sun can place constraints on the
SD WIMP-proton cross-section as collisions of WIMPs with hydrogen
(protons) is part of the process for capturing WIMPs in the Sun
\cite{Silk:1985ax}.
\Reffig{SDp} shows constraints placed by the IceCube/DeepCore
\cite{Aartsen:2012kia} and Baksan \cite{Boliev:2013ai} neutrino
detector searches for such neutrinos.  The constraints depend on the
annihilation channel and are shown here for the representative
$b$-quark and $W$-boson channels.  While neutrino searches can be very
sensitive to WIMPs with SD proton couplings, the high thresholds in
IceCube/DeepCore and some other neutrino experiments means they are
often unable to probe for light WIMPs as LUX and other direct searches
are capable of doing.  Furthermore, the limits shown here assume the
DM capture and annihilation processes in the Sun are in equilibrium,
an assumption that may not be true for many DM candidates \cite{Ellis:2009ka}.

The exclusion of the DAMA region by LUX in the SD proton-only coupling
case, the case where LUX limits are approximately at their weakest,
suggests that the LUX result may exclude \textit{any} SD explanation for
the DAMA signal, the first time a single experiment would be able to do so.
After a more careful examination over the mixed coupling case --- not
just the proton-only or neutron-only cases --- this is indeed the case:
the LUX likelihood-based limit at 90\%~CL excludes the entire SD
parameter space consistent with the DAMA result within the 2$\sigma$~CL,
at least for the assumed halo model.  As always, caveats apply.  Various
assumptions about detector behavior are made that, if incorrect, will
affect the interpretation of the experimental results and alter the WIMP
parameter space consistent with those results.  See for example
Ref.~\cite{Collar:2013gu}.  In addition, if the \NEST{} model for
low-energy events is inaccurate, the low-mass LUX limits may weaken.
For the more conservative maximum gap analysis, a tiny part of the
DAMA-compatible parameter space escapes the LUX bounds:
$\anSD/\apSD = -0.16\pm0.04$ with $\mchi\approx10$~GeV.  This remaining
space will be excluded by the next LUX data release if excess events are
not found.

\subsection{Application to effective theory}
\label{sec:EffectiveTheory}

We now apply the analysis of this paper to a DM model for which the SD constraints are particularly important. Consider the case of Dirac DM annihilating through the $s$-channel exchange of a spin-1 mediator, $V_\mu$, via an axial-vector interaction. Assuming that the mediator also has an axial-vector interaction with SM fermions, we obtain the Lagrangian:
\begin{equation}
\label{eqn:dmModelDirac}
\mathcal{L}\,\supset\,\left[\gxa\bar{\chi}\gamma^\mu\gamma^5\chi+\gfa\bar{f}\gamma^\mu\gamma^5f\right]V_\mu
\end{equation}
for DM that is a Dirac Fermion, and 
\begin{equation}
\label{eqn:dmModelMajorana}
\mathcal{L}\,\supset\,\left[\frac{1}{2}\gxa\bar{\chi}\gamma^\mu\gamma^5\chi+ \gfa\bar{f}\gamma^\mu\gamma^5f\right]V_\mu
\end{equation}
for Majorana DM, where $\gxa$ and $\gfa$ are unknown couplings~\cite{Berlin:2014tja,Lebedev:2014bba}. Assuming the limit of low-momentum exchange in the WIMP-nucleon scattering process, we can integrate out the mediator to obtain the following SD scattering cross-section for a Dirac WIMP:
\begin{equation} \label{eqn:effsigma}
\effsigma = \frac{4\mu^2 \gxa^2}{\pi m_v^4}J(J+1)\left[\frac{\Sp}{J}\,\effap+\frac{\Sn}{J}\,\effan\right]^2
\end{equation}
where $J$ is the spin of the nucleus, $m_v$ is the mediator mass and $\mu$ is the reduced WIMP-nuclear mass.  Here, $\effap$ and $\effan$ are the WIMP-proton and WIMP-neutron couplings, related to the couplings of \refsec{SD} via $\GNSD = 2\sqrt{2}G_F \aNSD = \gxa\effaN/m_v^2$.\footnote{Again, $\aNSD$ is sometimes used in the literature to refer to the $\GNSD$ normalization used here, as is the case with Ref.~\cite{Berlin:2014tja}.} For a Majorana WIMP, \refeqn{effsigma} adopts an additional factor of $1/2$. Though we focus here on direct searches for such a particle, colliders can also place constraints; see Ref.~\cite{Lebedev:2014bba} for a discussion.

Expressions for $\effap$ and $\effan$ can be derived by starting from the WIMP-quark interactions and performing a weighted sum over the components of each nucleon. Since the heavy quarks and gluons contribute negligibly to the spin content of the nucleon, DM scattering off nucleons is dominated by the sum over light quarks. This allows us to write: 
\begin{equation}
\tilde{a}_N=\sum_{q=u,d,s} \gfa\,\Delta_q^{(N)}
\end{equation}
where $\Delta_q^{(N)}$ is the nuclear spin content of quark $q$, and we have assumed that the coupling between the WIMP and each quark is identical and given by the coupling in our above Lagrangian. The standard values for the various $\Delta_q^{(N)}$ are $\Delta_u^{(\text{p})}=\Delta_d^{(\text{n})}=0.84$, $\Delta_u^{(\text{n})}=\Delta_d^{(\text{p})}=-0.43$ and  $\Delta_s^{(\text{p})}=\Delta_s^{(\text{n})}=-0.09$~\cite{Beringer:1900zz}.

Due to the symmetry in the up and down quark contributions, $\effan=\effap$, so $\anSD=\apSD$.
Neglecting the slight difference in the proton and neutron masses, $\effsigmap\approx\effsigman$
with
\begin{equation} \label{eqn:csFinal}
  \begin{split}
    \effsigman
      & = \frac{4\mu_{\chi\text{n}}^2 \gxa^2}{\pi m_v^4}
         \left(\frac{1}{2}\right) \left(\frac{3}{2}\right)
         \left[\frac{\frac{1}{2}}{\frac{1}{2}}\,\effan\right]^2 \\
      & = \frac{3\mu_{\chi\text{n}}^2 \gxa^2 \gfa^2}{\pi m_v^4}
          \Bigg[ \sum_{q=u,d,s} \Delta_q^{(\text{n})} \Bigg]^2
  \end{split}
\end{equation}
the WIMP-neutron cross-section, obtained using $J=\Sn=\frac{1}{2}$ and $\Sp=0$ in \refeqn{effsigma}, and $\effsigmap$ the WIMP-proton cross-section.

\begin{figure}[ht]
  \insertfig{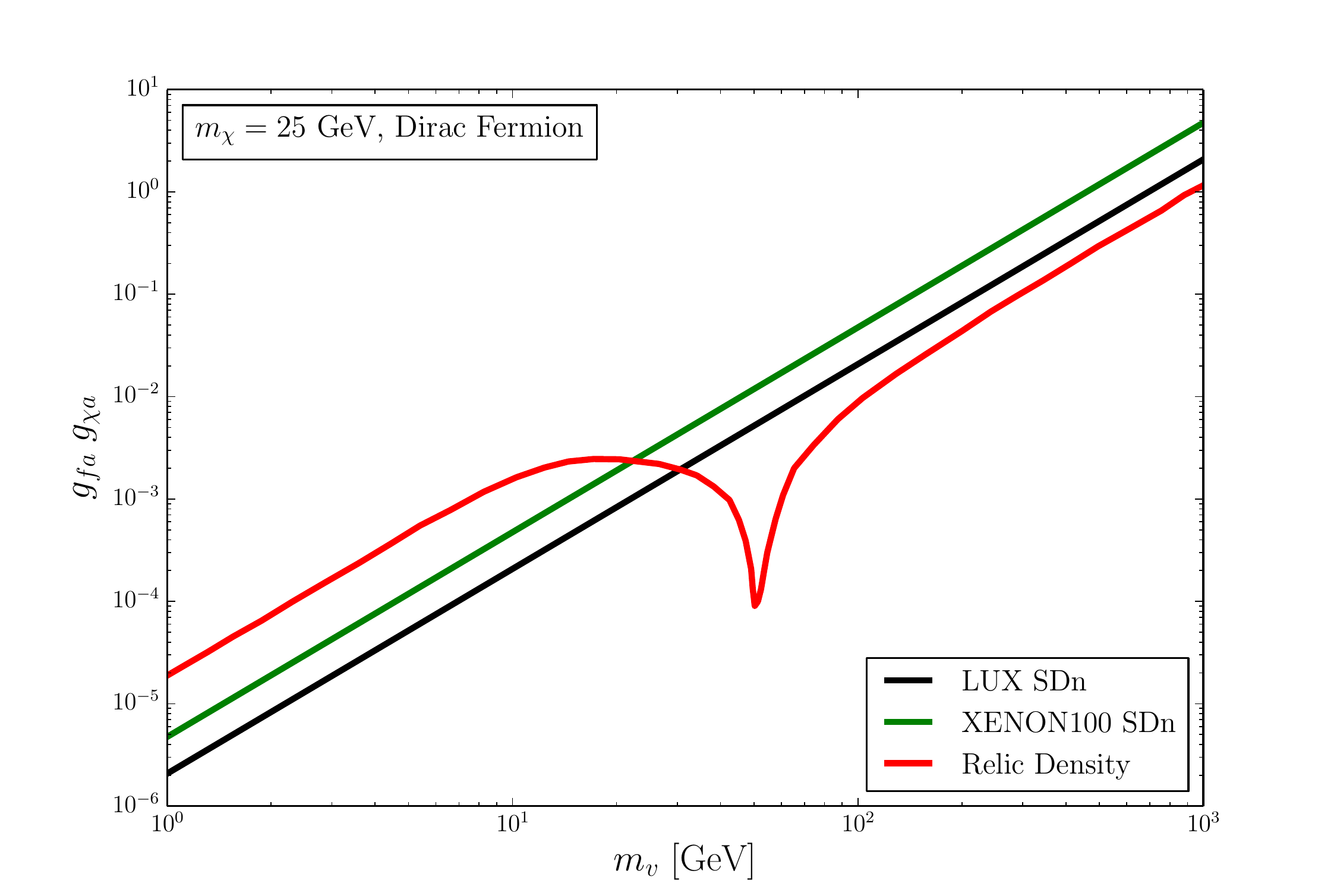}
  \caption{
    Constraints on the parameters of the effective Lagrangian given in \refeqn{dmModelDirac} for a 25~GeV Dirac WIMP. The LUX SDn limit obtained by \LUXCalc{} is shown in black, with coupling values above the black line excluded for a given mediator mass. Also shown are the results of a previous analysis utilising the XENON100 SDn limit (green line)~\cite{Berlin:2014tja} The red line shows the parameter values required to obtain the correct DM relic density ($\Omega_{\text{DM}}=0.268^{+0.013}_{-0.010}$) as measured by WMAP and Planck~\cite{Ade:2013zuv,Berlin:2014tja}.
    }
  \label{fig:dirac}
\end{figure}

\begin{figure}[ht]
  \insertfig{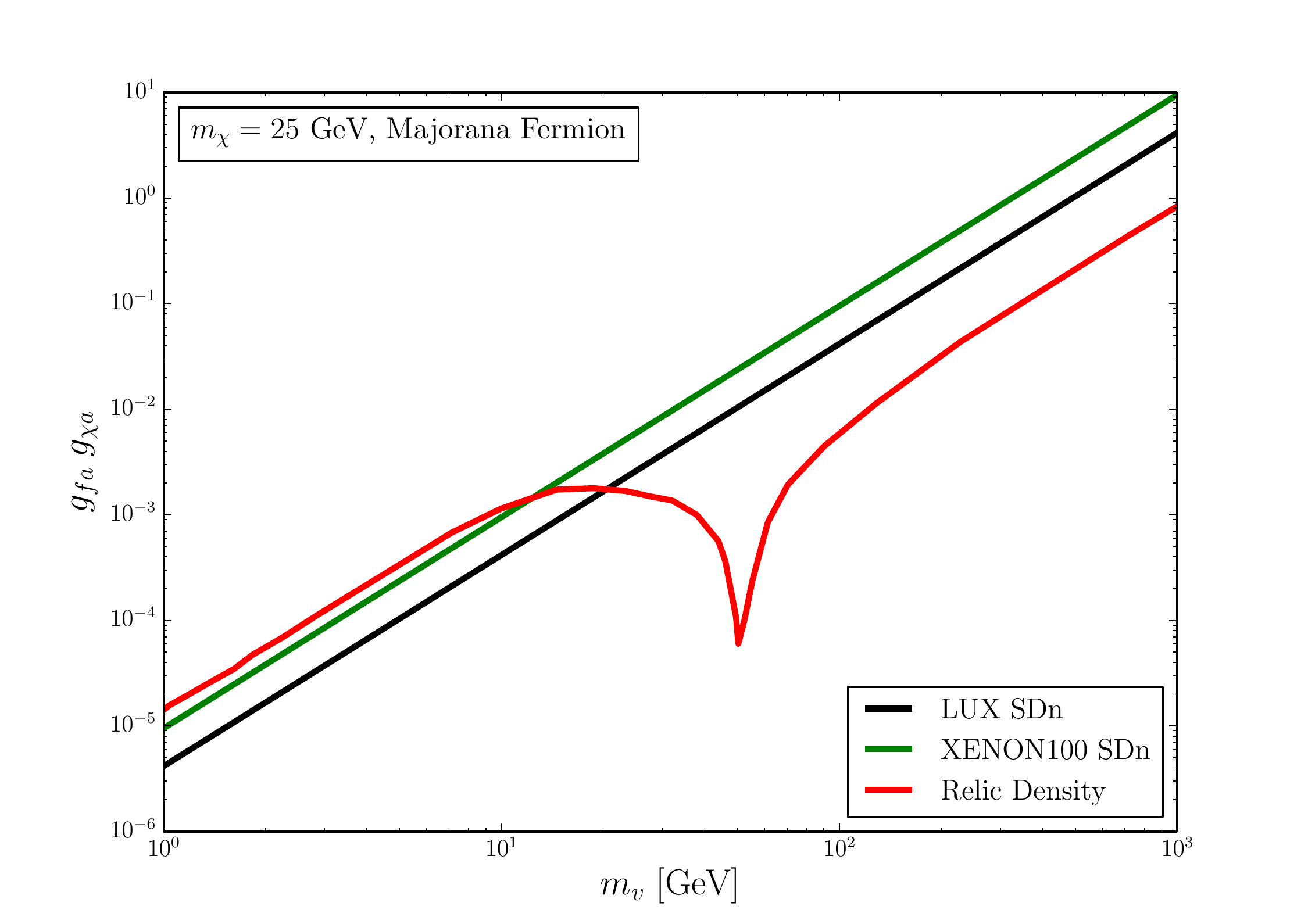}
  \caption{
    Similar to \reffig{dirac} but for a Majorana WIMP. 
    }
  \label{fig:maj}
\end{figure}


For a fixed WIMP mass, a limit on $\effsigman$ places a limit on $\gxa^2\gfa^2/m_v^4$, with that limit corresponding to a linear relationship between $\log(\gxa\gfa)$ and $\log{m_v}$, as shown in \reffig{dirac} (\reffig{maj}) for a Dirac (Majorana) WIMP. Here, we show LUX limits for a 25~GeV WIMP, using the 90\%~CL upper limit of $\effsigman < 1.26\times10^{-4}$~pb, determined by \LUXCalc{}. We stress a subtle point here: the appropriate LUX limit on $\effsigman$ is determined using the $\apSD=\anSD$ relation expected for this model, not the $\apSD=0$ assumed in the SD neutron-only case shown in \reffig{SDn}, though in practice the two limits are similar. We also show in Figures~\ref{fig:dirac} \&~\ref{fig:maj} the results of a previous analysis~\cite{Berlin:2014tja} based on the XENON100 result. As expected, the LUX bounds are more stringent, raising the limit on the mass of a mediator consistent with relic density observations (red curves) from 20~GeV to 30~GeV for a Dirac WIMP, and from 10~GeV to 20~GeV for a Majorana WIMP.

\section{Conclusions}
\label{sec:Conclusions}

The particle nature of DM is still unknown. There are several theoretical candidates for DM, however the WIMP hypothesis remains one of the most popular explanations of the phenomenon. The null results of direct detection experiments such as LUX provide limits on the physics of WIMP-nucleus interactions and therefore the parameter space of a given WIMP model. 

We have developed and utilized the new tool \LUXCalc{} to generate limits on the SI WIMP-nucleon cross-section at 90\%~CL both with Poisson and maximum gap based analyses for a WIMP mass in the range $[5,2500]$~GeV. We see that the maximum gap method generally agrees very well with the official LUX SI limits, whilst the Poisson likelihood-based constraint is weaker by $\sim 30\%$ above a WIMP mass of about 20~GeV. We then generate the LUX SI scattering limits for the low mass region $[3,15]$~GeV again using both the maximum gap and Poisson likelihood techniques. For the maximum gap method, we also show limits with the progressively more conservative exclusion of all contributions from events with energies below 1, 2, and 3~keV. These \LUXCalc{}-generated limits are then compared with the official LUX result as well as the anomalous signal regions as seen by CoGeNT, CRESST, and DAMA and the exclusion curve from SuperCDMS. In this mass region, the Poisson-likelihood curve provided the strongest constraint. 

We have used \LUXCalc{} to generate for the first time the LUX limits on the  SD WIMP-nucleon cross-section over the full range of WIMP masses. We show constraints for both neutron-only ($\apSD=0$) and proton-only ($\anSD=0$) couplings using the maximum gap method detailed in the text. We see that for the SD proton-only case, which is the SD case where the LUX limits are approximately at their weakest, the LUX limit fully excludes the DAMA region. In fact, we find that the LUX likelihood-based limit at 90\%~CL excludes the entire SD parameter space consistent with the DAMA result within the 2$\sigma$~CL (for the assumed parameters of the SHM). Furthermore, the more conservative maximum gap method excludes most of the DAMA parameter space, except for the small region $\anSD/\apSD=-0.16\pm 0.04$ with $\mchi \approx 10$~GeV.    

Finally, we have applied the main results of this work to an effective theory case where SD constraints on the WIMP-nucleon cross-section are particularly important. We see that the LUX bounds are more stringent than those of a previous study based on the XENON100 results \cite{Berlin:2014tja} by a factor of $\sim$2 in both the Dirac and Majorana WIMP cases.

We have made the \LUXCalc{} tool publicly available for future studies involving the LUX results.


\acknowledgments
  CS thanks P.~Sandick and the Department of Physics \& Astronomy at
  the University of Utah for support.
  CS thanks C.~Bal\'azs and the School of Physics at Monash University,
  where part of this work was performed, for their hospitality.
  CS was partially supported by NSF award 10681111.
  The work of AS and AGW is supported by the Australian Research Council
  through the Centre of Excellence for Particle Physics at the Terascale.
  MW is supported by the Australian Research Council Future Fellowship
  FT140100244.
  We thank J.~Cornell, P.~Scott, and M.~Szydagis for useful conversations. 



\onecolumngrid
\begingroup\leftskip0.075\paperwidth\rightskip0.075\paperwidth

\appendix

\section{LUXCalc}
\label{sec:LUXCalc}

Here we describe the \LUXCalc{} software package, which can be found at
Ref.~\cite{LUXCalc:url} or as ancillary files to the arXiv version of
this paper.  The package provides both a library and a standalone
program for performing various likelihood and constraint calculations.
The software is written in Fortran~95, but linking to the library can
be easily performed from \Cpp{}.
We begin in \refsec{LUXCalcRun} by describing some basic usage of the
standalone program, in \refsec{LUXCalcFortran} we show how to link to
the library from Fortran, and in \refsec{LUXCalcC} we show how to do
the same in \Cpp{}.  Finally, in \refsec{software}, we point out a few
routines in other software packages (\DarkSUSY{} and \micrOMEGAs{})
that may be useful.

Both the library and program can be compiled by running `\texttt{make~all}';
however, one of the \texttt{gfortran} or \texttt{ifort} compilers must
be installed.

\subsection{Program}
\label{sec:LUXCalcRun}

\lstset{style=Pstyle}

\begin{table*}
  \centering
  \addtolength{\tabcolsep}{0.5em}
  \begin{tabular}{lp{0.7\textwidth}}
    \hline 
    \texttt{--verbosity=1} &
      Specifies how much detail to provide in the output.  Increasing
      this above the default of 1 will cause a detailed header to be
      provided, as well as progressively more detailed data output.
      The additional data provided is mode specific.  \\
    \texttt{--rho=0.4} &
      The local density of the DM halo [GeV/cm$^3$].
      \\
    \texttt{--vrot=235.0} &
      The local rotation speed of the galactic disk [km/s].
      The velocity dispersion $v_0$ will be set to this value as
      expected for an isothermal spherical halo model (\ie\ the
      Standard Halo Model), unless specified via the \texttt{--v0}
      option.
      \\
    \texttt{--v0=235.0}   & \\
    \texttt{--vesc=550.0} &
      The velocity dispersion $v_0$ and local escape velocity $\vesc$
      (\ie\ cutoff speed) [km/s] used to define the truncated
      Maxwellian velocity distribution (\refeqn{TruncMaxwellian}).
      Specifically, the parameter $v_0$ is the most probable speed of
      the distribution in the absence of any truncation.
      \\
    \texttt{--Emin=0.0} &
      Only consider contributions to the expected signal from events
      with recoil energies above this value [keV].  Detector thresholds
      are already factored into the efficiencies, so contributions from
      low-energy events are suppressed regardless of this setting.
      \\
    \multicolumn{2}{l}{\texttt{--confidence-level=0.9}} \\
    \texttt{--p-value=0.1} &
      Specifies the confidence level (CL) or the p-value ($p=1-\mathrm{CL}$)
      to use for generating constraints.
      \\
    \multicolumn{2}{l}{\texttt{--m-tabulation=1.0,1000.0,-20}} \\
      &
      For quantities that are tabulated by mass like cross-section
      limits, this option specifies the minimum and maximum masses
      in the tabulation [GeV], followed by the number of tabulation
      intervals.  A negative number for the third value indicates
      number of intervals per decade. \\
    \hline 
  \end{tabular}
  \caption[Simulations]{%
    \label{tab:Options}%
    Useful options for running the \LUXCalc{} program.  Any values
    shown are the default values.  Some options are only useful in
    certain program modes.
    }
\end{table*}

The \LUXCalc{} program is called in the following form:
\begin{Psnippet}
  ./LUXCalc [mode] [options] [WIMP parameters]
\end{Psnippet}
where \texttt{[mode]} is a flag describing the type of calculation to
be performed, \texttt{[options]} are optional flags that can be used to
set various parameters or control the output, and \texttt{[WIMP parameters]}
are the WIMP mass and scattering cross-section(s), necessary only in
certain modes.  Several of the program modes are described below and
some of the most useful options are described in \reftab{Options}.
A full description of all modes and options can be found by running
`\texttt{./LUXCalc --help}'.

\medskip\noindent
\textbf{Likelihood.}
The logarithm of the Poisson-based likelihood, as described in
\refsec{LUX}, is given by
\begin{Psnippet}
  ./LUXCalc --log-likelihood [options] [WIMP parameters]
\end{Psnippet}
The WIMP parameters are a list of  values: the WIMP mass [GeV]
followed by one, two, or four cross-sections [pb]. In the first case,
the single cross-section is the SI WIMP-nucleon cross-section, assumed
to be the same for protons and neutrons.  In the second case, the two
cross-sections are the SI and SD WIMP-nucleon cross-sections, again
assuming identical couplings for protons and neutrons.  In the last
case, the four cross-sections are, in order, the SI WIMP-proton,
SI WIMP-neutron, SD WIMP-proton, and SD WIMP-neutron cross-sections.

\medskip\noindent
\textbf{Maximum gap p-value.}
The maximum gap p-value, as described in \refsec{LUX}, is given by
\begin{Psnippet}
  ./LUXCalc --log-pvalue [options] [WIMP parameters]
\end{Psnippet}
Specifically, this returns the quantity $1-C_0$, where $C_0$ is defined
in \refeqn{MaxGapC0} and is technically only an upper limit on the
p-value, not the p-value itself.  As opposed to the Poisson-based
likelihood above, this analysis involves no background subtraction and
is hence conservative.  The WIMP parameters are as described for the
likelihood mode above.

\medskip\noindent
\textbf{Likelihood constraints.}
Tabulated (by mass) upper and lower cross-section constraints (\ie\ 
confidence intervals) as determined via the likelihood are generated by
\begin{Psnippet}
  ./LUXCalc --constraints-SI [--theta-SI-pi=0.25] [options]
  ./LUXCalc --constraints-SD [--theta-SD-pi=0.25] [options]
\end{Psnippet}
where the two lines correspond to spin-independent (SI) and
spin-dependent (SD) interactions.  When determining constraints, the
ratio between WIMP-neutron and WIMP-proton couplings will be kept
fixed, with the ratio defined in terms of the polar angle $\theta$
in the $G_n$-$G_p$ plane, \ie\ $\tan\theta \equiv G_n/G_p = f_n/f_p = a_n/a_p$.
The default behavior is to take the two WIMP-nucleon couplings to be
identical; otherwise, $\theta$ can be specified (in units of $\pi$) via
the options as shown.  Increase the verbosity
(\eg\ \texttt{--verbosity=3}) to show the corresponding WIMP-neutron
constraints in addition to the WIMP-proton constraints.  The confidence
level (CL) of the confidence intervals is specified via the
\texttt{--confidence-level} option; the default is 90\%~CL.

\medskip\noindent
\textbf{Maximum gap limits.}
Tabulated (by mass) upper limits on the cross-section(s) as determined
by the maximum gap method are generated by
\begin{Psnippet}
  ./LUXCalc --limits-SI [--theta-SI-pi=0.25] [options]
  ./LUXCalc --limits-SD [--theta-SD-pi=0.25] [options]
\end{Psnippet}
where the options are as described for the likelihood constraints above.

\subsection{Library: Fortran usage}
\label{sec:LUXCalcFortran}

\lstset{style=Fstyle}

\LUXCalc{} is written as a single, self-contained Fortran~95 module.
All floating point values are in the \texttt{REAL*8} format, while
integers are of type \texttt{INTEGER}.
The module must be loaded in any user routine that calls
\LUXCalc{} routines:
\begin{Fsnippet}
  USE LUXCalc
\end{Fsnippet}

\medskip\noindent
\textbf{Initialization.}
Before calling any routines, the module must first be initialized with
\begin{Fsnippet}
  CALL LUXCalc_Init(intervals=.TRUE.)
\end{Fsnippet}
The single argument here specifies if calculations should be performed
for the intervals (gaps) between events.  This is necessary for
generating maximum gap limits, but is not required for any likelihood
calculations.  This initialization need only be performed once.
To force only recoils of energy greater than $\Emin$ to be considered
in calculating rates, use
\begin{Fsnippet}
  CALL LUXCalc_SetEmin(Emin=0d0)
\end{Fsnippet}
where the argument is in keV.  As noted elsewhere, detector thresholds
are already factored into the efficiencies, so contributions from
low-energy events are suppressed regardless of this setting.

\medskip\noindent
\textbf{Halo model.}
The parameters of the Standard Halo Model can be specified via
\begin{Fsnippet}
  CALL LUXCalc_SetSHM(rho=0.4d0,vrot=235d0,v0=235d0,  &
                      vesc=550d0)
\end{Fsnippet}
where the arguments are the local DM density $\rhochi$
[GeV/cm$^3$], the disk rotation speed $\vrot$ [km/s], the velocity
dispersion $\vmp$ [km/s], and the galactic escape speed $\vesc$ [km/s].
The values shown here are the defaults, which are already set when
\texttt{LUXCalc\_Init()} is called, so the above function call is
spurious.  The halo parameters can be modified at any time.

\medskip\noindent
\textbf{WIMP mass and couplings.}
The WIMP parameters are specified by one of three routines:
\begin{Fsnippet}
  CALL LUXCalc_SetWIMP_mfa(m,fp,fn,ap,an)
  CALL LUXCalc_SetWIMP_mG(m,GpSI,GnSI,GpSD,GnSD)
  CALL LUXCalc_SetWIMP_msigma(m,sigmapSI,sigmanSI,  &
                                sigmapSD,sigmanSD)
\end{Fsnippet}
Here, \texttt{m} is the WIMP mass $\mchi$ [GeV];
\texttt{fp} and \texttt{fn} are SI WIMP-proton and
WIMP-neutron couplings [GeV$^{-2}$], respectively;
\texttt{ap} and \texttt{an} are SD WIMP-proton and
WIMP-neutron couplings;
\texttt{GpSI}, \texttt{GnSI}, \texttt{GpSD}, and \texttt{GnSD} are
WIMP-nucleon couplings [GeV$^{-2}$], differing from $f$ and $a$ only in
normalization as discussed in \refsec{CrossSection}; and 
the \texttt{sigma} arguments are WIMP-nucleon cross-sections [pb].
A negative value for a cross-section can be used to indicate the
corresponding coupling should be taken to be negative.
The current mass, couplings, and cross-sections can be retrieved
with the corresponding routines
\begin{Fsnippet}
  CALL LUXCalc_GetWIMP_mfa(m,fp,fn,ap,an)
  CALL LUXCalc_GetWIMP_mG(m,GpSI,GnSI,GpSD,GnSD)
  CALL LUXCalc_GetWIMP_msigma(m,sigmapSI,sigmanSI,  &
                                sigmapSD,sigmanSD)
\end{Fsnippet}
The returned cross-section values will \textit{not} be set negative for
negative couplings.

\medskip\noindent
\textbf{Calculations.}
After any changes to the WIMP parameters and/or the halo distribution,
the LUX rate calculations must be performed using
\begin{Fsnippet}
  CALL LUXCalc_CalcRates()
\end{Fsnippet}
This routine performs the various LUX rate calculations that are used
for determining likelihoods and the various constraints. Thus, this
routine \textit{must} be called before obtaining expected events,
likelihoods, p-values, etc.  The relevant quantities are stored
internally.

\medskip\noindent
\textbf{Events.}
The number of observed and expected events for the current WIMP are
provided by the functions:
\begin{Fsnippet}
  INTEGER FUNCTION LUXCalc_Events()
  REAL*8 FUNCTION LUXCalc_Background()
  REAL*8 FUNCTION LUXCalc_Signal()
  REAL*8 FUNCTION LUXCalc_SignalSI()
  REAL*8 FUNCTION LUXCalc_SignalSD()
\end{Fsnippet}
In order, these return the observed number of events in LUX, the
average expected background events, the average expected signal events,
and the separate SI and SD contributions to the expected signal.

\medskip\noindent
\textbf{Likelihoods and p-values.}
The statistical functions for evaluating LUX results in the context of
the current WIMP are:
\begin{Fsnippet}
  REAL*8 FUNCTION LUXCalc_LogLikelihood()
  REAL*8 FUNCTION LUXCalc_LogPValue()
  REAL*8 FUNCTION LUXCalc_ScaleToPValue(logp)
\end{Fsnippet}
The first function returns the log of the likelihood using a Poisson
distribution in the number of observed events given the expected
background and signal.  The second function returns the logarithm of the
p-value in a more conservative no-background-subtraction analysis.
If \texttt{LUXCalc\_Init()} was called with a \texttt{.TRUE.} argument,
then the maximum gap method is used, with $p = 1-C_0$, where $C_0$ is
given by \refeqn{MaxGapC0}.  Otherwise, A Poisson distribution with zero
background contribution is assumed.  This function is only intended for
determining conservative one-sided limits as the value returned is
technically an upper limit on the p-value and not the p-value itself.
The last function determines such a limit by identifying the factor $x$
such that $\sigmas = x\sigmas_0$ gives the desired p-value
(specified as $\log(p)$), with $\sigmas_0$ the currently specified
WIMP-nucleon cross-sections.  The quantity $x\sigma_0$ is then the
limit on the cross-sections at the given confidence level ($1-p$),
assuming a fixed ratio of WIMP-nucleon couplings (\eg\ $\fpSI=\fnSI$).

\subsection{Library: C++ usage}
\label{sec:LUXCalcC}

To make usage of \LUXCalc{} with \Cpp\ code easier, a \Cpp\ interface
file \texttt{LUXCalc.hpp} is provided.  The routines and functions
are the same as those described for Fortran in the previous section,
with identical names and signatures, though Fortran subroutines
become \Cpp\ \texttt{void} functions.
All arguments and return values are \texttt{bool}, \texttt{int} or
\texttt{double}.

\subsection{Useful software}
\label{sec:software}

Here we show how to extract the relevant DM parameters from
two of the most popular software packages for examining DM in
the context of SUSY: \DarkSUSY{} \cite{Gondolo:2004sc,DarkSUSY:url} and
\micrOMEGAs{} \cite{Belanger:2001fz,Belanger:2008sj,Belanger:2013oya,MicrOMEGAs:url}.
\DarkSUSY{} is written in Fortran~77, with the various $G$ WIMP-nucleon
couplings for a given SUSY model provided by the \texttt{dsddgpgn}
routine. The WIMP (neutralino) mass must be retrieved from various
common blocks. The necessary parameters can be retrieved from
\DarkSUSY{} and set in \LUXCalc{} as shown in this minimal Fortran~95
example:
\begin{Fsnippet}
  ! Load LUXCalc module
  USE LUXCALC
  ! Variables to store WIMP mass and couplings
  REAL*8 :: M,GpSI,GnSI,GpSD,GnSD
  ! Calculated quantities (examples)
  REAL*8 :: lnlike,signal
  ! DarkSUSY common blocks defined in 'dsmssm.h'
  CHARACTER*8 :: pacodes_ctmp(0:50)
  INTEGER :: pacodes_itmp(60),kn(4),lsp,kln
  REAL*8 :: mass(0:50),mspctm_rtmp(6)
  COMMON /PACODES/ pacodes_itmp(1:18),kn,pacodes_itmp(23:60), &
                   pacodes_ctmp
  COMMON /MSPCTM/ mass,mspctm_rtmp
  COMMON /MSSMIUSEFUL/ lsp,kln
  ...
  ! Initialize LUXCalc
  CALL LUXCalc_Init(.FALSE.)
  ...
  ! For each SUSY model, do following >>>>>>>>
    ! Set WIMP parameters
    M = mass(kn(kln))
    CALL dsddgpgn(GpSI,GnSI,GpSD,GnSD)
    CALL LUXCalc_SetWIMP_mG(M,GpSI,GnSI,GpSD,GnSD)
    
    ! Calculate rates for current WIMP
    CALL LUXCalc_CalcRates()
    
    ! Get likelihood, expected signal events, etc.
    lnlike = LUXCalc_LogLikelihood()
    signal = LUXCalc_Signal()
    ...
\end{Fsnippet}
The \micrOMEGAs{} package provides the WIMP mass and couplings in the
\texttt{Mcdm} global variable and \texttt{nucleonAmplitudes} routine,
respectively; see Ref.~\cite{Belanger:2008sj} for a description of the
relevant \micrOMEGAs{} coupling routine and its arguments.
A minimal \Cpp{} example using \micrOMEGAs{} is:
\begin{Csnippet}
  // Initialize LUXCalc
  LUXCalc_Init(false);
  ...
  // For each SUSY model, do following >>>>>>>>
    // Set WIMP parameters
    // separate particle/anti-particle couplings
    double lambdap[2],lambdan[2],xip[2],xin[2];
    // FeScLoop is micrOMEGAs-provided function
    nucleonAmplitudes(FeScLoop,lambdap,xip,lambdan,xin);
    double M = Mcdm;  // Mcdm is global variable
    double GpSI = 2*lambdap[0];
    double GnSI = 2*lambdan[0];
    double GpSD = 2*xip[0];
    double GnSD = 2*xin[0];
    LUXCalc_SetWIMP_mG(M,GpSI,GnSI,GpSD,GnSD);
    
    // Calculate rates for current WIMP
    LUXCalc_CalcRates();
    
    // Get likelihood, expected signal events, etc.
    double lnlike = LUXCalc_LogLikelihood();
    double signal = LUXCalc_Signal();
    ...
\end{Csnippet}
We finally point out one subtlety: \DarkSUSY{} and \micrOMEGAs{} will
yield somewhat different WIMP-nucleon scattering cross-sections for a
given SUSY model due to different values chosen for the hadronic
matrix elements that go into calculating WIMP-nucleon couplings from
the WIMP-quark couplings \cite{Ellis:2008hf}. This is more of an issue
for the SI cross-sections, which typically differ by a factor of
$\orderof{2}$.  Both packages allow the matrix elements to be
modified; see their respective manuals for further details.

\endgroup
\twocolumngrid


\bibliography{LUXCalc}

\end{document}